\documentclass[12pt]{article}
\usepackage{graphicx}
\usepackage{amssymb}

\newcommand\pubnumber{CIPANP2015-Galati}
\newcommand\pubdate{\today}

\def\napoli{Universit\`a degli Studi di Napoli ``Federico II" and INFN}
\def\support{\footnote{on behalf of the OPERA Collaboration}}

\def\Title#1{\begin{center} {\Large #1 } \end{center}}
\def\Author#1{\begin{center}{ \sc #1} \end{center}}
\def\Address#1{\begin{center}{ \it #1} \end{center}}

\newcommand\pubblock{\rightline{\begin{tabular}{l} \pubnumber\\
         \pubdate  \end{tabular}}}
\newenvironment{Abstract}{\begin{quotation}  }{\end{quotation}}
\newenvironment{Presented}{\begin{quotation} \begin{center} 
             PRESENTED AT\end{center}\bigskip 
      \begin{center}\begin{large}}{\end{large}\end{center} \end{quotation}}

\def\beq{\begin{equation}}
\def\eeq#1{\label{#1}\end{equation}}
\def\eeqn{\end{equation}}

\def\beqa{\begin{eqnarray}}
\def\eeqa#1{\label{#1}\end{eqnarray}}
\def\eeqan{\end{eqnarray}}

\let\bar=\overbar

\def\Dslash{\not{\hbox{\kern-4pt $D$}}}
\def\dslash{\not{\hbox{\kern-2pt $\del$}}}

\def\msb{{\bar{\ssstyle M \kern -1pt S}}}

\begin{document}
\begin{titlepage}
\pubblock

\vfill
\Title{OPERA neutrino oscillation search: status and perspectives}
\vfill
\Author{Giuliana Galati\support}
\Address{\napoli}
\vfill
\begin{Abstract}
OPERA is a long-baseline experiment at the Gran Sasso laboratory (LNGS) designed to search for $\nu_\mu \rightarrow \nu_\tau$ oscillations in appearance mode. OPERA took data from 2008 to 2012 with the CNGS neutrino beam from CERN. The data analysis is ongoing, with the goal of establishing $\nu_\tau$ appearance with high significance and improving the sensitivity to the sterile neutrino search in the $\nu_\mu$ $\rightarrow$ $\nu_e$ appearance channel. Current results will be presented and perspectives discussed.
\end{Abstract}
\vfill
\begin{Presented}
 CIPANP2015
%Higher resolution figs at ??http://www.home.edu/~goode/CIHP/\\
\\Vail (Colorado), May 5 2015
\end{Presented}
%6 pages, LaTeX, 3 eps figures. 
\vfill
\end{titlepage}
\def\thefootnote{\fnsymbol{footnote}}
\setcounter{footnote}{0}

\section{Introduction}
The OPERA (Oscillation Project with Emulsion tRacking Apparatus) experiment \cite{opera1,opera2} has been designed to conclusively prove the existence of $\nu_\mu$ $\rightarrow$ $\nu_\tau$ oscillations in an almost pure $\nu_{\mu}$ beam. The direct appearance search is based on the detection of $\tau$ leptons produced in $\nu_\tau$ charged current interactions (CC).
The OPERA detector is placed in the underground Gran Sasso Laboratory (LNGS), 730~km away from the neutrino source, in the high energy long-baseline CERN to LNGS beam (CNGS) \cite{opera3,opera4}. The average neutrino energy is about 17~GeV. The $\bar{\nu}_{\mu}$ contamination is 2.1$\%$ in terms of interactions; the $\nu_{e}$ and $\bar{\nu}_{e}$ contaminations are in total below 1$\%$, while the number of prompt $\nu_{\tau}$ is negligible.

\section{The OPERA detector}
The challenge of the OPERA experiment is the detection of the short-lived $\tau$ lepton (c$\tau$ = 87~$\mu$m) produced in the CC $\nu_{\tau}$ interactions. Thus, one has to accomplish the very difficult task of detecting sub-millimeter $\tau$ decay topologies out of almost ten thousands of $\nu_{\mu}$ interactions in a target of 1.25 kilotons. This is achieved in OPERA using the nuclear emulsion technique that features an unrivaled spatial resolution ($\leq 1\mu m$).
\\The OPERA detector (10~m $\times$ 10~m $\times$ 20~m) is an hybrid apparatus that consists of an emulsion/lead target complemented by electronic detectors. The detector is made up of two identical super-modules aligned along the CNGS beam direction, each made of a target section and a muon spectrometer. Each target section consists of a multi-layer array of 28 target walls interleaved with pairs of planes of plastic scintillator strips. A target wall is an assembly of horizontal trays each loaded with Emulsion Cloud Chamber target units, called bricks. In the whole apparatus there are more than 150000 bricks. Each brick consists of 57 emulsion films, 300~$\mu m$ thick, interleaved with 56 lead plates, 1 mm thick. It has 128~mm $\times$ 102~mm $\times$ 70~mm outer dimensions and a mass of 8.3 kg. 
\\ The electronic detectors provide time resolution to the emulsions, preselect the interaction region, identify muons and measure their charge and momentum. Thanks to nuclear emulsions it is possible to study in detail interaction or decay topologies.

%Interface emulsion detectors, called Changeable Sheets, are attached to the downstream face of each brick.

\section{Event selection and analysis}
The electronic detectors are used to trigger the neutrino interactions and to locate the brick in which the interaction took place. Events are classified as charged current-like (1$\mu$) if there is a muon track reconstructed or a minimum amount of material is traversed, neutral-like (0$\mu$) otherwise \cite{mcs}.
Every brick containing an interaction is extracted from the walls asynchronously with respect to the beam to allow for film development, scanning and for the search for $\tau$ decays, which are carried out in the emulsion scanning laboratories in Europe and Japan with automated optical microscopes. Starting from a set of predictions provided by the electronic detectors, tracks of secondary particles produced in a neutrino interaction are followed back in the brick, film by film, from the most downstream one to the interaction point where they originate.
%Microscopes acquire a sequence of tomographic images of the emulsion layers, which are processed and analysed in order to recognise aligned clusters of dark pixels (grains) produced by charged particles along their trajectories, which form the tracks.
Whenever a track disappearance signal is detected (the track is not found in three consecutive films), a volume of $1$~cm$^2$ around the candidate vertex point is scanned in order to fully reconstruct the event and find any decay candidate through a dedicated decay search procedure.
% which consists of topological and kinematical selection criteria.
%In particular, computerize all tracks information for the later analysis.
If a secondary vertex is found, a full kinematical analysis is performed combining the measurements
in the nuclear emulsion with data from the electronic detectors. The momentum of charged
particles can be measured in emulsions by the Multiple Coulomb Scattering up to 6~GeV/c with resolution better than 22\% using the angular deviations \cite{mcs0}. It can be measured up to 12~GeV/c with
a resolution better than 33\% using position deviations. For muons crossing the spectrometers,
the momentum is measured with a resolution better than 22\% up to 30~GeV/c, the muon charge
is also determined \cite{mcs}. The hint of a decay topology is the observation of an impact parameter
larger than 10~$\mu$m, defined as the minimum distance between the track and the reconstructed
vertex, excluding low momentum tracks. The appearance of the $\tau$ lepton is identified by the detection of its characteristic decay topologies, either in one prong (electron, muon or hadron) or in three prongs. Kinematical selection criteria are then applied according to the decay channel.

\section{Search for $\nu_\mu$ $\rightarrow$ $\nu_\tau$ oscillations}
\begin{figure}[htb]
\centering
\includegraphics[height=2.5in]{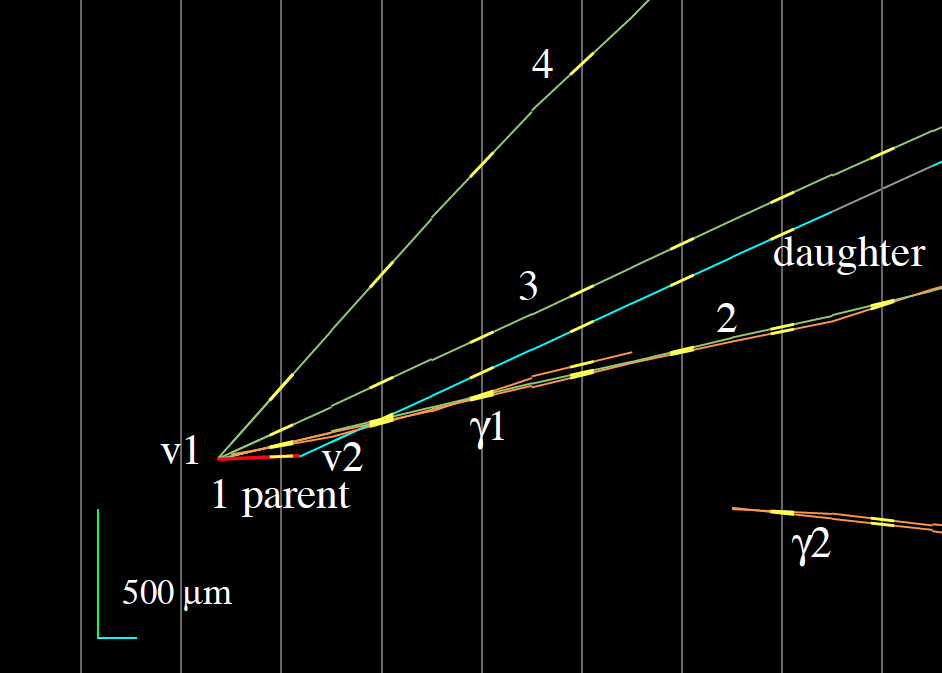}   
\caption{\label{fig:tau4} Event display of the fourth $\nu_\tau$ candidate in the projection longitudinal
to the neutrino direction. The primary and secondary vertices are indicated as $v1$ and $v2$, respectively.
}
\end{figure}

Runs with CNGS neutrinos were successfully carried out from 2008 to 2012. With a total CNGS beam intensity of $17.97 \cdot 10^{19}$ protons on target (p.o.t.), about 19505 neutrino events have been reconstructed by the electronic detector. About 7000 neutrino interactions have a vertex fully reconstructed in the emulsions. Out of them, 4685 events are used for the analysis reported in \cite{tau4}.
\\Four candidates $\tau$ events have been observed by the end of 2014, satisfying the kinematic selection criteria.
\\The first $\nu_\tau$ candidate was observed in the 2008-2009 data sample and described in detail in 2010 \cite{tau1}. 
It consists of a 7-prongs neutrino interaction. One of the tracks exhibits a kink topology and the daughter track is identified as a hadron through its interaction. Two $\gamma$-rays points to the secondary vertex. Their invariant mass (($120\pm20(stat.)\pm35(syst.))$~MeV/c$^2$) is compatible with the $\pi^0$ mass; their combination with the secondary hadron, assumed to be a $\pi^-$, gives an invariant mass of $640^{+125}_{-80} (stat.) ^{+100}_{-90} (syst.)$~MeV/c$^2$. 
For these reasons, the decay mode is compatible with $\tau\to\rho(770)\nu_\tau$, whose
branching ratio is about 25\%.\\
The second $\nu_\tau$ candidate was found in the 2011 data sample \cite{tau2}. The event has 2 prongs at primary vertex with the production of a short track with flight length of 1.54~mm (associated with $\tau$ lepton) and a longer track identified as a hadron. The $\tau$ lepton decays into 3 prongs, which are identified as hadrons on the basis of momentum-range consistency. 
All the kinematical cuts for the selection of $\tau\to3h$ decays
are satisfied.\\
The third  $\nu_\tau$ candidate was observed in the 2012 
%1$\mu$ 
data sample \cite{tau3}. 
The primary vertex is defined by two tracks and a $\gamma$-ray.
One of the tracks is identified as being a hadron based on momentum-range selection criteria. The other track is identified as the $\tau$ lepton decaying into a muon; the decay daughter matches with the muon track reconstructed by the electronic detectors, with a measured momentum of $(2.8 \pm 0.2)$~GeV/c and a negative
charge assessed with a $5.6~\sigma$ significance. The kinematical analysis of the event finally satisfies all the specified criteria for the $\tau \to \mu$ decay channel.\\
A new  $\nu_\tau$ candidate was found in 2012 
%0$\mu$ 
data sample \cite{tau4}.
The primary vertex is defined by four tracks and two $\gamma$-rays, as shown in Figure \ref{fig:tau4}. Track 1 is the parent track of a kink topology with an angle of $(137 \pm 4)$~mrad. The flight length is ($1090 \pm 30$)~$\mu$m.
The kink daughter track has a momentum of $(6.0^{+2.2}_{-1.2})$~GeV/c. Its impact parameter
with respect to the primary vertex is $(146 \pm 5)~\mu$m. The daughter track was followed in the downstream
bricks till the end of the target. It is exiting the target, stopping in the spectrometer
after leaving a signal in three RPC planes. Momentum-range correlation allows to identify it as a hadron.
\\

In the analysed sample 0.233 background events
are expected,
coming essentially from charmed events with an undetected primary muon,
hadronic re-interactions (for the hadronic decay channels) and large angle
muon scattering (for the $\tau \to \mu$ channel) as shown in detail in Table \ref{tab:sigbk}. Taking into account the different
signal-to-noise ratio for each decay channel, the four observed candidates
give a $4.2~\sigma$ significance for the exclusion of the background-only hypothesis of $\nu_\mu \to \nu_\tau$ oscillations.\\
Since the CIPANP 2015 Conference, a new $\tau$ candidate has been reported, allowing to assess the discovery of $\nu_{\mu}\rightarrow\nu_{\tau}$ oscillations in appearance mode with a significance of $5.1~\sigma$ \cite{tau5}.
%Given the 4 observed events and the expected background events, the confidence interval of $\Delta m^2_{23}$ is estimated with the Feldman-Cousins method \cite{fc}, assuming maximal mixing. The 90\% confidence interval of $\Delta m^2_{23}  is \left[1.8, 5.0\right] \times10^{-3}$ eV$^2$. An alternative analysis employing a Bayesian approach \cite{bay} with a flat  prior on $\Delta m^2_{23}$ was performed. In this case the credible interval of $\Delta m^2_{23}$  is $\left[1.9, 5.0\right] \times 10^{-3}$ eV$^2$, in agreement with the previous one.

\begin{table}[t]
\begin{center}
\begin{tabular}{l|ccccccc}
\hline
\multicolumn{1}{|c}{} & \multicolumn{1}{|c}{} & \multicolumn{1}{|c}{} & \multicolumn{4}{|c|}{Expected background} \\ 
\multicolumn{1}{|c}{{\small Decay}}  & \multicolumn{1}{|c}{{\small Expected}} & \multicolumn{1}{|c}{} & \multicolumn{1}{|c}{} &
\multicolumn{1}{|c}{{\footnotesize Charm}} & \multicolumn{1}{|c}{{\footnotesize Hadronic}} &
\multicolumn{1}{|c|}{{\footnotesize Large-angle}}  \\
\multicolumn{1}{|c}{{\small Channel}}  & \multicolumn{1}{|c}{{\small signal}} & \multicolumn{1}{|c}{{\footnotesize Observed}} & \multicolumn{1}{|c}{{\footnotesize Total}} &
\multicolumn{1}{|c}{{\footnotesize Decays}} & \multicolumn{1}{|c}{{\footnotesize re-interactions}} &
\multicolumn{1}{|c|}{{\footnotesize muon scattering}}  \\ \hline

\multicolumn{1}{|c}{{\small$\tau \rightarrow 1h$}} & \multicolumn{1}{|c}{{\footnotesize$0.41\pm 0.08$}} & \multicolumn{1}{|c}{{\small 2}} & \multicolumn{1}{|c}{{\footnotesize$0.033\pm 0.006$}} & \multicolumn{1}{|c}{{\footnotesize$0.015\pm 0.003$}} & \multicolumn{1}{|c}{{\footnotesize$0.018\pm 0.005$}} & \multicolumn{1}{|c|}{/} \\
\multicolumn{1}{|c}{{\small$\tau \rightarrow 3h$}} & \multicolumn{1}{|c}{{\footnotesize$0.57\pm 0.11$}} & \multicolumn{1}{|c}{{\small 1}} & \multicolumn{1}{|c}{{\footnotesize$0.155\pm 0.030$}} & \multicolumn{1}{|c}{{\footnotesize$0.152\pm 0.030$}} & \multicolumn{1}{|c}{{\footnotesize$0.002\pm 0.001$}} & \multicolumn{1}{|c|}{/} \\
\multicolumn{1}{|c}{{\small$\tau \rightarrow \mu$}}& \multicolumn{1}{|c}{{\footnotesize$0.52\pm 0.10$}} & \multicolumn{1}{|c}{{\small 1}} & \multicolumn{1}{|c}{{\footnotesize$0.018\pm 0.007$}} & \multicolumn{1}{|c}{{\footnotesize$0.003\pm 0.001$}} & \multicolumn{1}{|c}{/} & \multicolumn{1}{|c|}{{\footnotesize$0.014\pm 0.007$}} \\
\multicolumn{1}{|c}{{\small$\tau \rightarrow e$}} & \multicolumn{1}{|c}{{\footnotesize$0.62\pm 0.12$}} & \multicolumn{1}{|c}{{\small 0}} & \multicolumn{1}{|c}{{\footnotesize$0.027\pm 0.005$}} & \multicolumn{1}{|c}{{\footnotesize$0.027\pm 0.005$}} & \multicolumn{1}{|c}{/} &\multicolumn{1}{|c|}{/} \\

\hline

\multicolumn{1}{|c}{{\small Total}} & \multicolumn{1}{|c}{{\footnotesize$2.11\pm 0.42$}} & \multicolumn{1}{|c}{{\small 4}} & \multicolumn{1}{|c}{{\footnotesize$0.233\pm 0.041$}} & \multicolumn{1}{|c}{{\footnotesize$0.198\pm 0.040$}} & \multicolumn{1}{|c}{{\footnotesize$0.021 \pm 0.006$}} & \multicolumn{1}{|c|}{{\footnotesize$0.014\pm 0.007$}}\\

\hline

\end{tabular}
\caption{Expected signal and background in the analysed sample and the number of observed events}
\label{tab:sigbk}
\end{center}
\end{table}

\section{Sterile neutrino search via $\nu_\mu \to \nu_e$ oscillations}

A systematic search for $\nu_e$ events was performed with the 2008-2009 data sample, where 505 0$\mu$ events were located, corresponding to an integrated intensity of $5.25\cdot10^{19}~$ p.o.t. \cite{nue}. 19 $\nu_{e}$ candidate events have been observed. This number is compatible with the expected $\nu_e$ from the beam contamination ($19.8\pm2.8$).
The current result on the search for the three-flavour neutrino oscillation yields an upper limit $\sin^{2}2\theta_{13}< 0.44$ $(90\%$ C.L.).

OPERA limits the parameter space available for a non-standard $\nu_{e}$ appearance suggested by the results of the LSND and MiniBooNE experiments. It further constrains the still allowed region around $\Delta m^2_{new} = 5\cdot10^{-2}$~eV$^{2}$. 
%For large $\Delta m^2_{new}$ values, the $90\% C.L.$ upper limit on $\sin^{2}(2\theta_{new})$ reaches $7.2\cdot10^{-3}$.
A Bayesian approach has been used and the upper limit on $\sin^2 2\theta_{new}$ reaches the value $7.2 \cdot 10^{-3}$.
\\These results are going to be updated: about 50 $\nu_e$ candidates have been observed in the full data set. 
%The reconstructed energy resolution is improved when the calorimetric measurement in the Target Trackers will be complemented by following the hadron tracks and the electron showers in the downstream bricks. 
OPERA should then be able to access the parameter region comparable to its sensitivity below $\sin^{2} 2\theta_{new} = 5.0\cdot 10^{-3}$.

%\begin{figure}[htb]
%\centering
%\includegraphics[height=2.5in]{nue3.pdf}
%\caption{\label{fig:nue3} The exclusion plot for the parameters of the non-standard $\nu_\mu \to \nu_e$ oscillation, obtained from this analysis using the Bayesian method, is shown. Limits from other experiments are also shown.
%}
%\end{figure}

\section{Sterile neutrino search via $\nu_\mu\rightarrow\nu_\tau$ oscillations}

\begin{figure}[htb]
\centering
\includegraphics[height=2.5in]{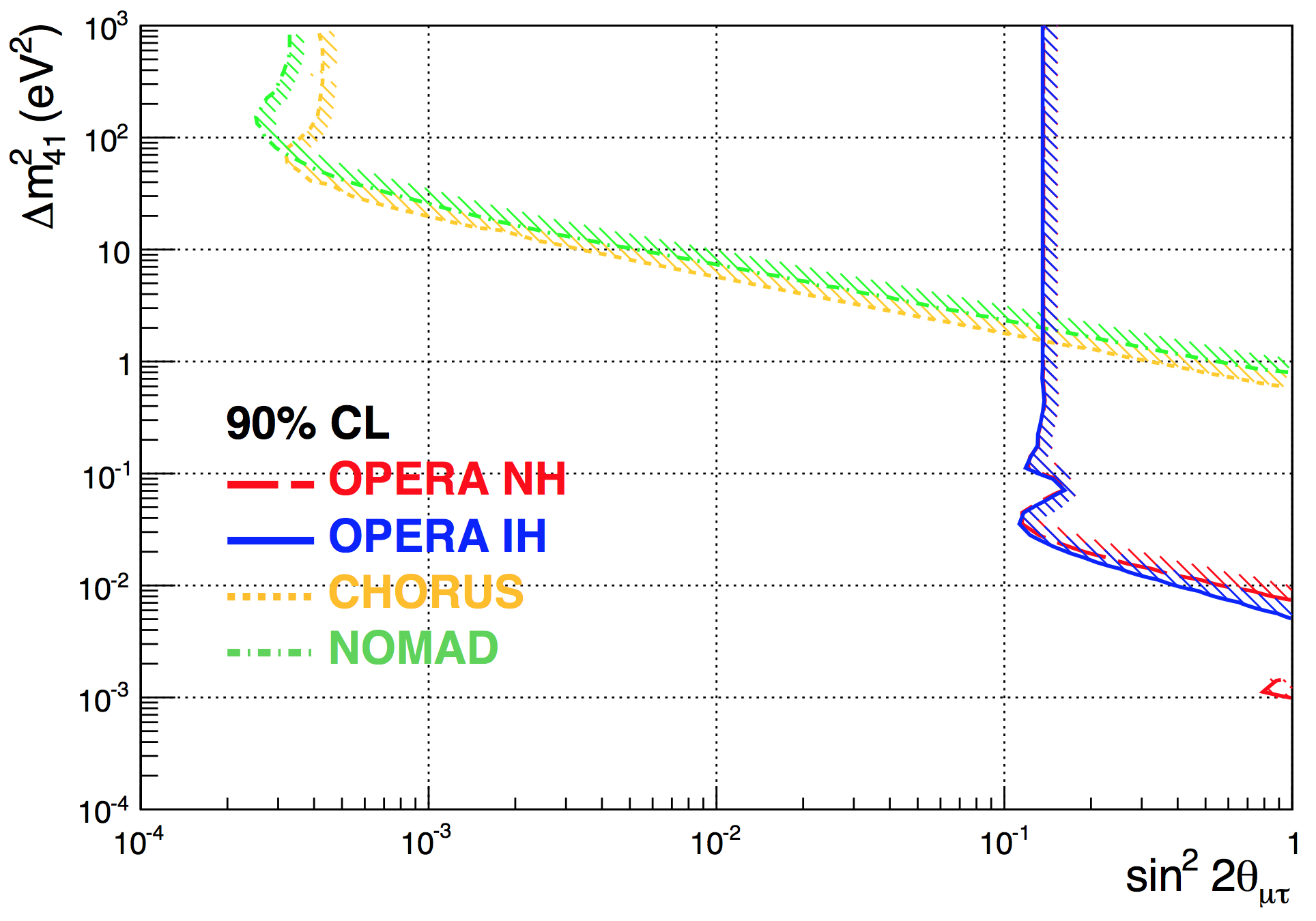}
\caption{\label{fig:ster} OPERA $90\%$~C.L exclusion limits in the $\Delta m^2_{41}$  vs $\sin^{2}2\theta_{\mu\tau}$  parameter space for the normal (NH, dashed red) and inverted (IH, solid blue) hierarchy of the three standard neutrino masses. The exclusion plots by other experiments are also shown. Bands are drawn to indicate the excluded regions.
}
\end{figure}

OPERA $\nu_\tau$ appearance results have been used to derive limits on the mixing parameters of a massive sterile neutrino \cite{sterile}.
\\In presence of a fourth sterile neutrino with mass $m_4$, the oscillation probability is a function of the $4\times 4$ mixing matrix $U$ and of the three squared mass differences.
Observed neutrino oscillation anomalies, if interpreted in terms of one additional sterile neutrino, suggest $|\Delta m^2_{41}|$ values at the eV$^{2}$ scale. In the framework of the 3+1 model, at high values of $\Delta m^2_{41}$, the measured $90\%$~C.L. upper limit on the mixing term $\sin^{2}2\theta_{\mu\tau} = 4|U_{\mu4}|^{2}|U_{\tau4}|^{2}$ is 0.116, independently of the mass hierarchy of the three standard neutrinos. The OPERA experiment extends the exclusion limits on $\Delta m^{2}_{41}$ in the $\nu_{\mu}\rightarrow\nu_{\tau}$ appearance channel down to values of $10^{-2}$eV$^{2}$ at large mixing for $\sin^{2}2\theta_{\mu\tau}\gtrsim 0.5$ as shown in Figure \ref{fig:ster}.

\section{Conclusions}
The OPERA experiment has been taking data from 2008 to 2012, collecting \nobreak{$17.97 \cdot 10^{19}$ p.o.t}.; the analysis at the emulsion
level is still on going.
\\Four $\nu_\tau$ candidates have been observed by the end of 2014 and the non-null observation of $\nu_\mu \to \nu_\tau$ oscillations is excluded at $4.2~\sigma$. In June 2015 a new $\tau$ candidate has been reported, achieving the discovery of $\tau$ neutrino appearance with $5.1~\sigma$ significance \cite{tau5}.
\\The observed number of $\nu_e$ interactions is compatible with the non-oscillation hypothesis, allowing OPERA to set an upper limit in the parameter space for a non-standard $\nu_e$ appearance.
\\Limits on the mixing parameters of a massive sterile neutrino have also been derived and the exclusion limits on $\Delta m^{2}_{41}$ in the $\nu_{\mu}\rightarrow\nu_{\tau}$ appearance channel has been extended down to values of $10^{-2}$ eV$^{2}$ at large mixing for $\sin^{2}2\theta_{\mu\tau}\gtrsim 0.5$.

%%%%%%%%%%%%%%%%%%%%%%%%%%%%%%%%%%%%%%%%%%%%%%%%%%%%%%%%%%%%%%%%%%%%%%%%%
%%
%%   use this format to include an .eps figure into your paper
%%
%\begin{figure}[htb]
%\centering
%\includegraphics[height=1.5in]{magnet}
%\caption{Plan of the magnet used in the mesmeric studies.}
%\label{fig:magnet}
%\end{figure}
%%%%%%%%%%%%%%%%%%%%%%%%%%%%%%%%%%%%%%%%%%%%%%%%%%%%%%%%%%%%%%%%%%%%%%%%%%%

%%%%%%%%%%%%%%%%%%%%%%%%%%%%%%%%%%%%%%%%%%%%%%%%%%%%%%%%%%%%%%%%%%%%%%%%%
%%
%%   use this format to include a LaTeX table  into your paper
%%
%\begin{table}[t]
%\begin{center}
%\begin{tabular}{l|ccc}  
%Patient &  Initial level($\mu$g/cc) &  w. Magnet &  
%w. Magnet and Sound \\ \hline
% Guglielmo B.  &   0.12     &     0.10      &     0.001  \\
% Ferrando di N. &  0.15     &     0.11      &  $< 0.0005$ \\ \hline
%\end{tabular}
%\caption{Blood cyanide levels for the two patients.}
%\label{tab:blood}
%\end{center}
%\end{table}
%%%%%%%%%%%%%%%%%%%%%%%%%%%%%%%%%%%%%%%%%%%%%%%%%%%%%%%%%%%%%%%%%%%%%%%%%%%

%\Acknowledgements

\bibliographystyle{unsrt}

\end{document}